\documentclass[12pt]{article}

\newcommand{\be}{\begin{equation}}
\newcommand{\ee}{\end{equation}}

\begin{document}
\begin{center}
\Large{\bf {Coherent States for the Non-Linear Harmonic Oscillator}}\\
\vskip 1cm
Subir Ghosh\\
\vskip .5cm
Physics and Applied Mathematics Unit\\
 Indian Statistical Institute\\
 203 B. T. Road, Kolkata 700108, India
\end{center}
\vskip .3cm
{\bf {Abstract:}} 
Wave packets for the Quantum Non-Linear Oscillator are considered in the
Generalized Coherent State
framerwork. To first order in the non-linearity parameter the Coherent State
behaves very similarly
to its classical counterpart. The position expectation value oscillates in a simple
harmonic manner. The energy-momentum
uncertainty relation is time independent as in a harmonic oscillator. Various
features, (such as the Squeezed State nature), of the Coherent
State have been discussed.

\vskip .5cm

In this paper we apply a recently developed scheme \cite{sp} of constructing
Generalized Coherent States (GCS) \cite{kl},\cite{gcs} to a  widely studied model:
Quantum
Non-linear Harmonic Oscillator (QNHO) \cite{ml}, with  interesting
consequences.  Its classical analysis reveals periodic solutions. One can exploit its shape invariance property to generate exactly
the energy spectrum and eigen-functions \cite{car}. It has also been analyzed 
as a Harmonic Oscillator (HO) with position dependent mass \cite{br}. In certain
limits similar models have appeared \cite{sg1} in oscillator models compatible
with a non-commutative
$\kappa$-Minkowski spacetime. Quite interestingy, it was shown that although the
coordinate
undergoes a ``simple'' harmonic motion, actually the full Hamiltonian operator
appears in the equation of motion in place of the frequency that appears in
conventional harmonic
oscillator.

Our scheme is computationally very simple. It is perturbative in nature. We
restrict ourselves to $O(\lambda )$ results where for $\lambda =0$ the
Non-linear HO reduces to simple HO. We will exploit the factorization property
\cite{car} to derive a Darboux-like transformation to rewrite  the QNHO in terms
of canonical creation-annihilation operators in a HO Fock basis. We will
explicitly demonstrate that the GCS behave in a very interesting manner and the
classical behaviour is qualitatively preserved. At various stages we will
compare and contrast features of QNHO with
the GCS \cite{sp} constructed for Non-Commutative HO compatible with the
Generalized Uncertainty Principle \cite{gup}

{\bf{I. 1-D Quantum Non-linear Harmonic Oscillator:}} The one-dimensional
Lagrangean model of the QNHO is \cite{ml,car,br}
\begin{equation}
L=\frac{1}{2}(\frac{1}{1+\lambda x^2})(\dot x^2-\alpha^2x^2).
\label{l}
\end{equation}
For $\lambda =0$ we get back HO. With $$p=(\partial L)/(\partial \dot x)=\dot
x/(1+\lambda x^2)$$ one obtains the Hamiltonian
\begin{equation}
H=p\dot x-L=\frac{1}{2}((1+\lambda x^2)p^2+\frac{\alpha^2x^2}{1+\lambda x^2}).
\label{h1}
\end{equation}
It has been shown \cite{car} that the  quantum Hamiltonian operator
corresponding to (\ref{h1}) admits a factorization $H'=H-\beta /2~,~H'=A^\dagger
A$ where,
\begin{equation}
A=\frac{1}{\sqrt{2}}(i{\sqrt{1-\lambda x^2}}p+\frac{\beta}{{\sqrt{1-\lambda
x^2}}}x),$$$$
A^\dagger=\frac{1}{\sqrt{2}}(-i{\sqrt{1-\lambda
x^2}}p+\frac{\beta}{{\sqrt{1-\lambda x^2}}}x)
\label{A}
\end{equation}
provided
\begin{equation}
\alpha^2=\beta(\beta +\lambda ).
\label{al}
\end{equation}
The energy eigenvalues are,
\begin{equation}
<n\mid A^\dagger A\mid n>=\beta n-\frac{\lambda }{2}n^2,
\label{en}
\end{equation}
\begin{equation}
\frac{1}{2}<n\mid A^\dagger A +AA^\dagger \mid n>=\beta
(n+\frac{1}{2})-\frac{\lambda }{2}n^2.
\label{een}
\end{equation}
In the above $n$ is an integer.  We
drop the zero point energy and consider the energy to be,
\begin{equation}
<n\mid A^\dagger A\mid n>=\beta n-\frac{\lambda }{2}n^2.
\label{en1}
\end{equation}
$\lambda $ can be both positive and negative. Hence for $\lambda \le 0$ $n$ is
unrestricted but for $\lambda \ge 0$ the allowed integer values of  $n$ are
restricted by
$n\leq (2\beta )/\lambda $.\\
{\bf{II. Canonical map of $x,p$ in terms of $a,a^\dagger $:}} It will be
convenient for our 
purpose to express $x,p$ in terms of canonical creation-annihilation operators
$a,a^\dagger $.
For $\lambda =0,~A_{\lambda =0}=\sqrt \beta a,~A^\dagger_{\lambda =0}=\beta a
a^\dagger$ where $a$
is the canonical annihilation operator written in terms $x,p$ which can in turn
be
inverted to express $x,p$ in terms of $a,a^\dagger $. In the present case we
wish to do the same for non-zero $\lambda$: express   $x,p$ in terms of
$a,a^\dagger $ to $O(\lambda
)$.
Since
\begin{equation}
[\frac{1}{\sqrt 2}(ip+\beta x),\frac{1}{\sqrt 2}(-ip+\beta x)]=\beta
\label{xp}
\end{equation}
we have
\begin{equation}
a\equiv \frac{1}{\sqrt {2\beta}}(ip+\beta x)~,~~a^\dagger \equiv \frac{1}{\sqrt
{2\beta }}(-ip+\beta x)
\label{a}
\end{equation}
and furthermore
\begin{equation}
x=\frac{1}{\sqrt {2\beta}}(a+a^\dagger
)~,~~p=-i\sqrt{\frac{\beta}{2}}(a-a^\dagger ).
\label{xxp}
\end{equation}
Quite clearly the above constitute the $\lambda =0$ relations. Now to $O(\lambda
)$
\begin{equation}
A\approx \frac{1}{\sqrt{2}}(i{\sqrt{1-\lambda x^2}}p+\beta x(1-\frac{\lambda
}{2})$$$$
\approx\frac{1}{\sqrt{2}}((ip+\beta x)+\frac{\lambda }{2}x^2(ip-\beta x)),
\label{A1}
\end{equation}
\begin{equation}
A^\dagger 
\approx\frac{1}{\sqrt{2}}((-ip+\beta x)+\frac{\lambda }{2}x^2(-ip-\beta x)),
\label{A1}
\end{equation}
Now we need to be careful since operator ordering is involved{\footnote{For a
rigorous and mathematical discussion on the issue of operator ordering in the
present problem see \cite{car}.
We, on the other hand, take a more naive approach, one reason being that we are
considering only $O(\lambda )$
extension and this might clash with the exact analysis provided in \cite{car}.}}
. We take care of
this below when we write $A,A^\dagger $ in terms of $a,a^\dagger $ and invoke
Weyl ordering.
\begin{equation}
A={\sqrt {\beta }}(a-\frac{\lambda }{4\beta}(a+a^\dagger)^2a^\dagger
)_{WO}~,$$$$
A^\dagger ={\sqrt {\beta }}(a^\dagger -\frac{\lambda }{4\beta}(a+a^\dagger)^2a
)_{WO}
\label{AA}
\end{equation}
From the combination using the exact relations (\ref{al}),
\begin{equation}
(A+A^\dagger )=\frac{\sqrt {2}\beta x}{{\sqrt{1+\lambda x^2}}}
\label{Ax}
\end{equation}
we obtain
\begin{equation}
x=\frac{1}{\sqrt 2\beta}((A+A^\dagger )+\frac{\lambda }{4\beta }(a+a^\dagger
)^3)_{WO} $$$$
=\frac{1}{\sqrt {2\beta}}(a+a^\dagger ).
\label{xa}
\end{equation}
This simple algebra leads us to the cherished expressions,
\begin{equation}
x=\frac{1}{\sqrt {2\beta}}(a+a^\dagger
)~,~~p=-i\sqrt{\frac{\beta}{2}}(a-a^\dagger ).
\label{xxp}
\end{equation}
It is somewhat unexpected to find out that {\it{to the first non-trivial order
in}}
$\lambda$, $x,p$ {\it{retain their canonical form when expressed in terms of}}
$a,a^\dagger$. The HO Fock space is
\begin{equation}
a\mid n>={\sqrt{n}}\mid n-1>~,~~a^\dagger\mid n>={\sqrt{n+1}}\mid n+1>.
\label{n}
\end{equation}
From here on the computations are straightforward. The Hamiltonian to $O(\lambda
)$ is obtained from $H=A^\dagger A+\frac{\beta}{2}$ and (\ref{AA}) with the
necessary 
Weyl ordering. In HO Fock space
representation we find
\begin{equation}
 H=\frac{1}{2}[\beta (aa^\dagger +a^\dagger
a)+\frac{\lambda}{2}\{a^2+(a^\dagger)^2+aa^\dagger +a^\dagger a )$$$$-(a^4+
(a^\dagger)^4+a^2(a^\dagger)^2+(a^\dagger)^2a^2
+\frac{(a^\dagger)3a}{2}+\frac{
a^\dagger a^3}{2}+\frac{a(a^\dagger)^3}{2}+\frac{a^3a^\dagger}{2}
$$$$+\frac{a^\dagger a(a^\dagger)^2}{2}+\frac{a^2a^\dagger
a}{2}+\frac{(a^\dagger)^2aa^\dagger}{2}
+\frac{aa^\dagger a^2}{2} )\}]+\frac{\beta}{2}.
\label{ha}
\end{equation}
Since we are considering first order perturbation theory only terms with equal
number of $a$ and $a^\dagger $ will contribute to the energy expectation value
in state $\mid n>$,
\begin{equation}
 H\mid n>=(\beta n-\frac{\lambda}{2}n^2+\frac{1}{2}(2\beta
-\frac{\lambda}{2})\mid n>.
\label{hn}
\end{equation}
One can compare the above energy expression with the exact value \cite{car},
with (\ref{en},\ref{een}). Note that the $n$-dependent
terms in the exact energy
(\ref{een}) and our first order corrected value (\ref{hn}) are identical. In
fact for our GCS construction the constant shift in energy is unimportant and
henceforth will be ignored.

{\bf{III. Generalized Coherent States for NQHM:}} After these preliminaries we
are now ready for the main piece of our work:
construction of the GCS. We follow the notation of \cite{gcs} and the GCS $\mid
J,\gamma >$ is the following wave packet,
\begin{equation}
\mid J,\gamma >=\frac{1}{N^2(J)}\sum_{n\geq 1}\frac{J^{n/2}e^{-i\gamma
e_n}}{{\sqrt{\rho_n}}}\mid n>~,~~\rho_n =e_1e_2...e_n .
\label{J}
\end{equation}
The parameter $\gamma $ is proportional to $\beta $ and $J$ is related to $<H>$
for the GCS.
In the above $e_n$ is defined as 
\begin{equation}
E_n=\beta n(1-\frac{\lambda}{2\beta}n)=\beta
n(1-\frac{\lambda '}{2}n)=\beta e_n,~~ \lambda '=\frac{\lambda }{\beta}.
\label{en1}
\end{equation}

{\bf{IV. Properties of the Generalized Coherent States for NQHM:}} 
We start with the Revival time analysis. Recalling the energy expression as
$E_n=\beta n-(\lambda /2) n^2$ we find that there are two time scales involved:
the Classical time
$T_{c}=(2\pi)/ \beta $ and the Revival time $T_r=(4\pi )/\lambda $ with the
condition that $\lambda /(2\beta )$
an integer. Hence the motion with period $T_c$ will be modulated by $T_r$. Since
we consider $\lambda $ to be small, $T_r\>>T_c$. Full revival of the wave packet
will occur at each
multiple or $T_r$. In between $T_r$ there will be fractional revival where the
wave packet collapses into
subsidiary packets that evolve with period $T_c$. This fraction revival
phenomena and the Revival time scale
is a manifestation of the non-linearity in the system, showing up in the
non-linear energy spectrum.

Next we come to the features of the GCS. The building
blocks for 
further analysis is $<a>,<a^\dagger >$, expectation values of $a,a^\dagger $ in
the GCS. We find
\begin{equation}
<a>\equiv < J,\gamma \mid a\mid J,\gamma >=\frac{1}{N^2(J)}\sum_{n\geq
1}\frac{J^{n-\frac{1}{2}}e^{-i\gamma (e_{n-1}-e_n)}}{{\sqrt{\rho_{n-1}\rho_n}}}
\label{<a>}
\end{equation}
 The phase turns out to be
\begin{equation}
e_{n-1}-e_n=-(n+\frac{\lambda '}{2}(1-2n)),
\label{ee}
\end{equation}
and the exponential can be expanded as a power series in $\lambda '$. Hence we
find
\begin{equation}
<a>\approx \frac{\sqrt J}{N^2}e^{i\gamma}\sum_{n\geq
1}\frac{J^n(1-\frac{n}{2}(1+2n))}{\rho_n(1-\frac{\lambda ' n}{2})}.
\label{aap}
\end{equation}
A simple algebra leads to,
\begin{equation}
<a>={\sqrt {J}}e^{i\gamma}(1-\frac{\lambda '}{2}(1+J))~,~~<a^\dagger >={\sqrt
{J}}e^{-i\gamma}(1-\frac{\lambda '}{2}(1+J)).
\label{aj}
\end{equation}
This immediately yields the cherished expressions for the quantum behaviour of
position and momentum in coherent states,
\begin{equation}
 <x>={\sqrt{\frac{2J}{\beta}}}(1-\frac{\lambda '}{2}(1+J))cos\gamma ,
\label{x}
\end{equation}
\begin{equation}
 <p>={\sqrt{2J\beta}}(1-\frac{\lambda '}{2}(1+J))sin\gamma .
\label{p}
\end{equation}
It is very interesting to note that, for the GCS, non-linearity affects only the
amplitude, keeping the  oscillatory motion intact. This is one of our major
observations.

At this point it is worthwhile to compare this feature of NQHO with GCS of
another recently studied 
HO with non-linear deformation \cite{sp}. The latter system is an extension of
HO in a noncanonical phase
space that is compatible with the Generalized Uncertainty Principle \cite{gup}.
For the latter system, compared to (\ref{en1})
\begin{equation}
 e^{GUP}_n \approx n(1+\lambda (1+n)),
\label{gn}
\end{equation}
for which 
\begin{equation}
<x> \approx [cos \gamma -\lambda
((1+\frac{J}{2})cos\gamma +2(1+J)\gamma sin\gamma )],
\label{j1}
\end{equation}
\begin{equation}
<p> \approx [-sin\gamma 
+\lambda (\frac{1}{6}(2+J)sin\gamma -2\gamma (1+J)cos\gamma +\frac{J}{3}sin
(3\gamma ))].
\label{j3}
\end{equation}
Notice that for the GUP HO the time dependence is much more involved with higher
frequencies coming in to play.

Following the same procedure, although more complicated, one can compute the
dispersions,
\begin{equation}
 (\Delta x)^2=<x^2>-(<x>)^2=\frac{1}{\beta}[\frac{1}{2}+\lambda '
J(\frac{3}{2}(1+J)$$$$+(\frac{7}{4}+\frac{3}{2}J)cos(2\gamma) +2(1+J)\gamma
sin(2\gamma))],
\label{dx}
\end{equation}

\begin{equation}
 (\Delta p)^2=<p^2>-(<p>)^2=\beta[\frac{1}{2}+\lambda '
J(\frac{3}{2}(1+J)$$$$-(\frac{7}{4}+\frac{3}{2}J)cos(2\gamma) -2(1+J)\gamma
sin(2\gamma))].
\label{dp}
\end{equation}
Here we find a marked qualitative difference from the HO behaviour since the
non-linearity introduces a $\gamma $ or time-dependent oscillatory motion.
{\it{However it is remarkable that this time dependence disappears in the
uncertainty
relation to give}}
\begin{equation}
 (\Delta x)^2(\Delta p)^2=[\frac{1}{4}+\frac{3}{2}\lambda ' J(1+J)].
\label{dxp}
\end{equation}
This is another of our interesting observations. We point out that for the GUP
HO \cite{sp} the time
independent behavior of the Uncertainty Product of $x$ and $p$ is not
maintained.

It is worthwhile to point out that the GCS is an example of a Squeezed State.
Consider the instant
$\gamma =0$ in the oscillating variances $(\Delta x)^2$ and $(\Delta p)^2$ in
(\ref{dx}),(\ref{dp}),
 \begin{equation}
 (\Delta x)^2=\frac{1}{\beta}[\frac{1}{2}+\lambda '
J(\frac{13}{2}+3J)]~, ~
(\Delta p)^2=\beta[\frac{1}{2}-\frac{1}{4}\lambda '
J].
\label{sq}
\end{equation}
Clearly $(\Delta p)^2\leq\beta[\frac{1}{2}$ showing that the GCS is a Squeezed
State and that 
$(\Delta p)^2$ and $(\Delta x)^2$ attains their minimum and maximum values
respectively. The opposite happens for
$\gamma =\pi /2$. However it is not a minimum uncertainty Squeezed State since
$(\Delta x)^2(\Delta p)^2\geq\frac{1}{4}$.

The Weyl ordered Hamiltonian yields the GCS energy expectation value,
\begin{equation}
 <H>=\frac{\beta}{2}(1+2J)+\lambda '
[\frac{\beta}{2}J(1+J)+\frac{J^2}{4}-\frac{1}{4}(1+J+2Jcos(2\gamma ))^2].
\label{h}
\end{equation}
An interesting point is to note that for non-zero $\lambda ' $, $<H>$ has a $cos
(2\gamma )$ dependence
indicating that there is a little fuzziness in energy of the packet. Since
$\gamma \sim \beta $ the $cos$-term averages out for
$\beta t >>1$. From the condition (\ref{al}) 
\begin{equation}
\beta ^2+\beta\lambda ' -\alpha ^2=0~\rightarrow \beta \approx (\lambda \pm
2\alpha
)/2,
\label{eav}
\end{equation}
we find the oscillatory behavior can be ignored for $t>>\beta ^{-1}\approx
\alpha^{-1}(1+\frac{\lambda '}{2}\alpha ^{-1})$.

A direct way to ascertain the non-classical behavior is to construct the Mandel 
parameter $Q$ out of the dispersion in number operator,
\begin{equation}
Q=(\Delta n)^2/<n> -1
\label{q0}
\end{equation}
where $(\Delta n)^2=<N^2>-<N^2>$. For the present problem we find,
\begin{equation}
<N>=<a^\dagger a>=J[1+\frac{\lambda ' }{2}(1+J)],
\label{n}
\end{equation}
\begin{equation}
<N^2>=<a^\dagger aa^\dagger a >=J[1+J+\lambda ' (\frac{1}{2}+1+J^2)],
\label{n^2}
\end{equation}
leading to
\begin{equation}
(\Delta n)^2=<N^2>-<N^2>=J+2\lambda ' J(1+J)[-\frac{1}{2}+J+J^2].
\label{dn2}
\end{equation}
The Mandel parameter follows,
\begin{equation}
Q=\frac{\lambda ' }{2}(1+J)(4J^2+3J-3).
\label{q}
\end{equation}
One denotes $Q=0$ as  the Poissionian statistics and $Q\ge 0 $ ($Q\leq 0
$) as Super-Poissionian  (Sub-Poissionian) statistics. The distribution will be 
Super-Poissionian ($Q\geq 0$) for $J\geq 0.5$ and Sub-Poissionian ($Q\leq
0$) for $0.5\geq J\geq 0$. Poission statistics is recovere for
$J=(\sqrt{57}-3)/8$.

The remaining task is to check up on the status of the Correspondence Principle.
First we discuss the quantum  equation of motion by directly
 exploiting the Heisenberg
equation of motion,
\begin{equation}
 \dot B =i[H,B]~\rightarrow ~<\dot B> =i<[H,B]>,
\label{dot}
\end{equation}
for a generic observable $B$. In the present case  we obtain the following
operator equations,
\begin{equation}
 \dot a=-\frac{i}{2}[2\beta a +\lambda (a^\dagger +a-2(a^\dagger
)^3-a^2a^\dagger-a^\dagger a^2$$$$-(a^\dagger )^2a
-a^3-a(a^\dagger )^2-a^\dagger a a^\dagger )],
\label{dota}
\end{equation}
\begin{equation}
 \ddot a=-\beta ^2a+\lambda \beta (-a+a^2a^\dagger+a^\dagger
a^2-2(a^\dagger)^3+2a^3 ).
\label{ddota}
\end{equation}
Expectation values of the above along with their hermitian conjugates lead to
the equation of motion
\begin{equation}
 <\ddot x>=\frac{1}{{\sqrt{2\beta }}}<\ddot a+\ddot (a^\dagger
)>={\sqrt{2J\beta}}[-\beta +\lambda ' (-1+2J+(1+J)\frac{\beta}{2})]cos\gamma .
\label{xx}
\end{equation}
Let us now consider the classical equation of motion. The Hamiltonian equations
of motion yield,
\begin{equation}
 \dot x=p+\lambda x^2p~,~~\dot p=-\alpha^2x-\lambda (p^2x-2\alpha^2x^3) $$$$
\rightarrow ~\ddot x=-\alpha^2x+2\lambda x\frac{p^2+\alpha
^2x^2}{2}=-\alpha^2x+2\lambda xH +O(\lambda ^2).
\label{xcl}
\end{equation}
Now we compute the expectation value  keeping in mind the
operating ordering,
\begin{equation}
 <\ddot x>=<(-\beta ^2 x+\lambda (-\beta x+4\frac{xH+Hx}{2}))>$$$$
={\sqrt{2J\beta}}[-\beta +\lambda ' (-1+(4+\frac{\beta}{2})(1+J)]cos\gamma.
\label{xxcl}
\end{equation}
One immediately observes the striking similarity between the equation of motion
obtained from Heisenberg (quantum) and
Hamiltonian (classical) formalisms. \\
{\bf{V. Conclusion:}} We have constructed wave packets as Generalized Coherent
States for the Quantum
Non-Linear Harmonic Oscillator to the first non-trivial order in the
non-linearity parameter. Quite
remarkably the wave packet  closely mimics the behavior of position variable in
classical simple harmonic oscillator.
The oscillatory motion of the Coherent State in the quantum Non-linear
oscillator only has a modified amplitude. The uncertainty relation for the
Coherent State is time independent with only its magnitude modified by
non-linearity parameter. Once again the similarity  with quantum harmonic
oscillator
striking. Furthermore the Coherent State has the charecteristics of a Squeezed
State.
 The Correspondence Principle is also very nearly
maintained. Mandel parameter
analysis shows that  departures from the Poissionian behavior is possible
depending on Coherent State parameters and
Poission statistics is recovered for a particular value of the parameter.\\
{\bf{ Acknowledgement:}} I thank Bikashkali Midya for discussions.

\end{document}